\documentclass[fleqn,usenatbib]{mnras}

\usepackage{newtxtext,newtxmath}

\usepackage{graphicx}
\usepackage{subcaption}
\usepackage[T1]{fontenc}

\DeclareRobustCommand{\VAN}[3]{#2}
\let\VANthebibliography\thebibliography
\def\thebibliography{\DeclareRobustCommand{\VAN}[3]{##3}\VANthebibliography}

\usepackage{amsmath}	

\usepackage{orcidlink}

\title[Bar Formation in MaNGA Galaxies.]{Bar Formation in Disc Galaxies: Internal Kinematics and Environmental Influence in MaNGA Galaxies.}

\author[Erik Aquino-Ortiz et al.]{Erik Aquino-Ortíz$^{1}$\orcidlink{0000-0003-1083-9208} \thanks{E-mail: e.aquino@irya.unam.mx}, Bernardo Cervantes-Sodi$^{1}$\orcidlink{0000-0002-2897-9121} and Karol Chim-Ramirez$^{1}$\orcidlink{0009-0009-7473-9864}
\\
$^{1}$ Instituto de Radioastronomía y Astrofísica, Universidad Nacional Autónoma de México, Antigua Carretera a Pátzcuaro 8701,\\
Ex-Hda. San José de la Huerta, 58089 Morelia, Michoacán, Mexico}

\date{Accepted 2025 December 18. Received 2025 December 11; in original form 2025 July 7}

\pubyear{\the\year{}}

\begin{document}
\label{firstpage}
\pagerange{\pageref{firstpage}--\pageref{lastpage}}
\maketitle

\begin{abstract}

We explore how the physical properties of disc galaxies relate to the presence of bars using data from the SDSS-IV MaNGA survey. By combining internal kinematical properties and environmental diagnostics, we find that barred galaxies are more frequently associated with centrally concentrated stellar mass distributions (within 1 and 2 effective radii) and exhibit lower values of the stellar angular momentum $\lambda_{R_e}$.  At fixed total stellar mass, barred galaxies exhibit: \textit{(i)} higher stellar mass, and \textit{(ii)} lower angular momentum, both in their inner regions than their unbarred counterparts. We find a bimodal dependence of the bar fraction on tidal interactions produced by the nearest neighbour. Specifically, the bar fraction peaks in the most isolated galaxies, where bars form unequivocally through internal secular processes, decreases at intermediate interaction strengths, and rises again in the strong interaction regime, likely reflecting the role of dense environments in sustaining or triggering bars. Our results suggest that internal gravitational instabilities are the primary driver of bar formation. External tidal perturbations play a secondary role, capable of triggering or enhancing bar formation in galaxies that are already internally predisposed. Our findings provide robust observational validation of theoretical bar formation and evolution models in galaxies.
\end{abstract}




\section{Introduction}

About $50-70\%$ of disc galaxies show a stellar bar when observed in near–infrared bands, which are less affected by dust and better trace the underlying stellar mass distribution \citep[e.g.][]{Eskridge2000,Mendez-Delmestre2007,Erwin2018}. In optical wavelengths, the global bar fraction is somewhat lower, with $\sim30\%$ typically corresponding to strong bars only \citep[e.g.][]{Sellwood&Wilkilson1993,Aguerri2009,Masters2011,Toñito2022MNRAS,Chacon2024,Wang2025}. Bar formation in spiral galaxies is generally attributed to two main mechanisms: while theoretical studies traditionally attribute bar formation to internal dynamical instabilities in self-gravitating, axisymmetric discs \citep[e.g.,][]{Ostriker1973ApJ, Athanassoula2013Review, Sellwood2014, Lokas2019a}, modern simulations emphasize that gravitational interactions, including flyby encounters, mergers, or tidal forces from galaxy clusters can externally induce bars \citep[e.g.,][]{Miwa1998, Gauthier2006ApJ, Lokas2014, Lokas2016, MartinezValpuesta2017, Peschken2019, Semczuk2024}. These interactions can trigger the onset of bar instabilities by disturbing the disc equilibrium, or enhancing non-axisymmetric features. However, depending on the timing, geometry, and strength of the interaction, such perturbations can also delay or suppress bar formation by heating the disc or disrupting the conditions required for bar growth \citep[e.g.,][]{Lokas2016,Moetazedian2017,Lokas2018,Zana2018,Ghosh2021}. Regarding internal factors, several physical conditions are known to play a critical role in determining the susceptibility of spiral galaxies to bar formation. One of the most relevant is the relative dominance of the stellar disc over the dark matter halo in the inner regions of galaxies. Early theoretical studies showed that a massive, rigid halo can stabilize discs against bar formation by reducing their susceptibility to non-axisymmetric instabilities \citep{Ostriker1973ApJ}. When the halo dominates the central gravitational potential, it suppresses bar growth or delays its onset \citep[e.g.,][]{Mayer2004, Fujii2018}. Conversely, in systems where the stellar disc dominates the inner gravitational potential—so-called “maximal” discs—bars form more readily. Under these conditions, the disc becomes sufficiently self-gravitating to develop non-axisymmetric structures \citep[e.g.,][]{Debattista2000}. The dominance of the stellar disc is typically quantified by the ratio between the stellar and total rotation curves at 2.2 scale lengths ($R_{\rm d}$), where the stellar contribution peaks \citep[$0.75 \leq V_{\star}/V_{\rm tot} \leq 0.95$;][]{Sackett1997}. Observational studies also support this scenario: galaxies with high halo-to-stellar mass ratios tend to exhibit lower bar fractions, whereas those with more massive, self-gravitating discs show a higher incidence of bars \citep[e.g.,][]{Cervantes-Sodi2015, Díaz-Garcia2016A&A587A_160D, Cervantes-Sodi2017}. In this sense, the classical criterion of \citet{Efstathiou1982} (ELN) provides a useful first-order approximation, encapsulating the intuition that bar formation is favoured in baryon-dominated systems, even though many other factors, such as velocity dispersion, gas content, and halo response, are also critical \citep[e.g.,][]{Athanassoula2008, Sellwood2014,Romeo2023}. Besides the barion-to-halo ratio, numerical experiments have shown that the angular momentum of the configuration affects the formation and evolution of stellar bars. \citet{Long2014ApJ} demonstrated that the growth of stellar bars in spinning dark matter halos is heavily suppressed in the secular phase and that a dark matter halo can emit and not purely absorb angular momentum, hindering the growth of bars. This dependence on the spin of the dark matter halo has also been found in self-consistent cosmological simulations \citep[e.g.,][]{Ansar2024,Chim2025} and confirmed in observational studies \citep{CervantesSodi2013, Cervantes-Sodi_Sanchez2017}. This duality between baryonic dominance and angular momentum highlights the complex interplay between internal disc structure and stability. Both factors can influence whether a galaxy develops and sustains a bar, while external perturbations or environmental effects may further modulate bar formation efficiency \citep[e.g.,][]{Zana2018, Lokas2019a, Zheng2025}. \\
Beyond their formation mechanisms, bars are recognized as key agents of secular evolution reshaping galactic structure in disc galaxies. Theoretically, bars are known to facilitate angular momentum transfer through resonant interactions within the disc, particularly between the bar and the surrounding stars and dark matter halo \citep[e.g.,][]{Lynden-Bell1972,Sellwood1980,Athanassoula2002, Debattista2006}. During their secular evolution, bars increase in length with time \citep[][]{Athanassoula_y_Misiriotis2002,Ghosh_y_DiMatteo2024}. Stellar bars can drive gas within them to the central regions of the galaxy, triggering central star formation \citep[e.g.,][]{Kormendy2004, Kormendy2013, Athanassoula2013, SanchezGarcia2023}. Due to vertical instabilities in the bar, these structures drive the formation of boxy/peanut-shaped bulges \citep[e.g.,][]{Combes1990A&A, Fragkoudi2015}. Spectroscopic observations from the ARGOS survey \citep{Ness2013} revealed that boxy/peanut bulges exhibit characteristic kinematic patterns, matching predictions from N-body simulations of bar buckling \citep{Martinez-Valpuesta2006}. This vertical instability redistributes both stars and gas in the inner regions \citep[e.g.,][]{Debattista2020}, modifying the gravitational potential and fueling subsequent secular evolution. Recent integral field spectroscopy studies with unprecedented spatial resolution have confirmed theoretical predictions regarding the impact of bars on spiral galaxies. These observations reveal the presence of boxy/peanut-shaped bulges through their distinct kinematic signatures, providing strong evidence that barred spiral galaxies in the local Universe exhibit clear signs of secular reshaping driven by bars \citep{Gadotti2020}. Previous work has demonstrated the influence of bars on galaxy kinematics. \citet{Seidel2015} systematically quantified the kinematic impact of bars in nearby galaxies, revealing clear imprints on their stellar angular momentum $\lambda_R$ profiles. Similarly, \citet{delMoral-Castro2020} reported significant differences in $\lambda_R$ between barred and unbarred spirals. These studies confirm that variations in $\lambda_R$ induced by bars encode meaningful physical information on angular momentum redistribution and secular evolution. \\ 
In this context, our goal is to observationally investigate how internal and environmental properties relate to the presence of bars in spiral galaxies. We achieve this by analysing the bar occurrence fraction across multidimensional parameter spaces, combining structural, kinematic, and environmental diagnostics from the SDSS-IV MaNGA survey \citep{Bundy2015_MaNGA}. We focus on three fundamental properties: \textit{(i)} the total stellar mass ($M_{\star}$), \textit{(ii)} the stellar-to-dynamical mass ratio ($M_{\star}/M_{\rm dyn}$) measured within one and two effective radii ($R_{\rm e}$), and \textit{(iii)} the stellar angular momentum $\lambda_{\rm R_e}$, originally introduced by \citet{Emsellem2007}. By comparing the stellar-to-dynamical mass ratio and the stellar angular momentum at fixed stellar mass between barred and unbarred galaxies, we aim to evaluate whether they follow distinct evolutionary pathways. Finally, we examine how the bar incidence varies as a function of the tidal strength exerted by the first nearest neighbour. This is characterized by the $Q_{nn}$ parameter, which has been precomputed and made publicly available by \citet{Argudo_Fernandez2015} as part of their Value Added Catalog\footnote{GEMA-VAC: Galaxy Environment for MaNGA Value Added Catalog: \url{https://www.sdss4.org/dr17/data_access/value-added-catalogs/?vac_id=gema-vac:-galaxy-environment-for-manga-value-added-catalog}}. The $Q_{nn}$ parameter is defined as:

\begin{equation}
    Q_{\rm nn}=log\left(\frac{M_{\rm neigh}}{M_{\rm gal}} \left(\frac{D_{\rm gal}}{d_{\rm nn}} \right)^{3} \right),
\end{equation}
where $M_{\rm gal}$ and $D_{\rm gal}$ refer to the stellar mass and estimated diameter of the primary galaxy, respectively, while $M_{\rm neigh}$ and $d_{\rm nn}$ correspond to the stellar mass and projected distance of its first nearest neighbour, this parameter quantifies the strength of external tidal perturbations, providing a complementary perspective to our analysis of the internal structure ($M_{\star}/M_{\rm dyn}$, $\lambda_{\rm Re}$). By examining the interplay between tidal forces and internal dynamics, we investigate how these interactions correlate with the presence of bars in galaxies with different structural and kinematic properties. The paper is organized as follows. In Section \ref{sec: data}, we briefly describe the data set. Section \ref{sec: Analysis} outlines the main analysis procedures, including the sample selection and the kinematic analysis used to derive the key physical parameters. The main results and discussion are presented in Section \ref{sec:Results}. Finally, we conclude with a summary of our findings in Section \ref{sec:Conclusions}.

\section{Data and Sample.}
\label{sec: data}

This study utilizes data from the Mapping Nearby Galaxies at Apache Point Observatory survey \citep[MaNGA;][]{Bundy2015_MaNGA}, one of the three projects included in the fourth generation of the Sloan Digital Sky Survey \citep[SDSS-IV;][]{Blanton2017AJ}.

\subsection{The MaNGA survey}

Between April 2014 and August 2020, the MaNGA survey utilized integral field spectroscopy (IFS) across a broad wavelength range, from 3,600 to 10,300 \AA\ at a spectral resolution of $R \sim 2000$, to observe more than 10,000 galaxies selected from the NASA-Sloan Atlas (NSA) within a redshift range of $0.01<z<0.15$, providing spatially resolved spectroscopic data for a representative sample of the nearby universe. The sample was selected to achieve a flat distribution of stellar mass with $M_{\star} \ge 10^{8.5} M_{\odot}$ and to ensure the observation of enough galaxies across all morphological types and environments \citep[][]{Bundy2015_MaNGA}. The observations were conducted using the SDSS 2.5-meter telescope at Apache Point Observatory \citep[][]{Gunn2006AJ} and the SDSS-III Baryon Oscillation Spectroscopic Survey (BOSS) spectrograph \citep[][]{Smee2013AJ}. MaNGA uses 17 integral field units (IFUs), grouping them into hexagonal bundles of optical fibers of different sizes, each containing 19 to 127 optical fibers with a diameter of 2 arcsec \citep[][]{Drory2015AJ}. Due to the gaps between the hexagonal bundles, the observations were dithered, adopting a three-point triangular pattern on the sky to achieve complete spatial coverage of the sources \citep[][]{Law2015AJ}. The final MaNGA sample contains three main sub-samples: \textit{(i)} the primary subsample, comprising 60\% of the galaxies, selected so that the fields of view cover a $1.5$ galaxy effective radius ($r_e$); \textit{(ii)} the secondary subsample which comprises the 30\% of the galaxies, such that the fields of view cover at least $2.5\ R_e$, and $(iii)$ the color-enhanced subsample, which represents 10\% of the main sample, which includes low luminosity red galaxies, high luminosity blue galaxies, and green valley galaxies to fill in the poorly sampled regions of the color–magnitude diagram. For further details about the sample design see \citet{Wake2017AJ}. 

The MaNGA Data Reduction Pipeline \citep[DRP,][]{LawDRP2016AJ} reduces the single fibers in each exposure to produce flux-calibrated, sky-subtracted individual spectra from each exposure for a given galaxy. The result of the data reduction is a single data cube, in which the x-axis and y-axis correspond to the spatial coordinates (R.A. and DEC.) and the z-axis to the spectral information. Each channel in the z-axis corresponds to a monochromatic image, and each pixel at location (x,y) comprises a single spectrum. The final DRP data cubes were analyzed with the new implementation of the Pipe3D pipeline \citep[][]{SanchezPIPE3D2016RMxAA}, the pyPipe3D \citep[][]{Lacerda2022NewA}, which was developed to analyze spectra from IFS data. It is well known that the surface brightness of galaxies decreases with galactocentric distance, leading to a corresponding decline in the signal-to-noise ratio (S/N) in the outer regions. As a result, any analysis of the stellar continuum in low-S/N areas would be unreliable. Therefore, pyPipe3D applies spatial binning, aggregating adjacent spaxels to enhance the S/N, to a maximum goal of 50 in each spaxel, without significantly compromising the galaxy spatial resolution. Once the binning is done, the pipeline performs a multi-SSP decomposition of the stellar population and corrects for the binning effect afterward by adopting a dezonification procedure \citep[][]{CidFernandez2013A&A}. This procedure consists of re-scaling the best-fitting stellar population model, derived for each bin, to the flux intensity in each spaxel within the considered bin. As a result, all spectra within each bin have similar stellar population properties. This way, the pyPipe3D tool can deliver spatial maps of any stellar property, including the kinematics. 

Along this study, for the MaNGA data set, we use the following galaxy properties: \textit{(i)} the publicly available line-of-sight kinematic maps and stellar mass surface density maps derived by the pyPipe3D pipeline\footnote{pyPipe3D pipeline: \url{https://ifs.astroscu.unam.mx/pyPipe3D/}} \citep[][]{Sanchez2022}, \textit{(ii)} the effective radius, $R_e$, and total stellar mass, $M_{\star}$, extracted from the NSA catalog (\citet{Blanton2011}, http://www.nsatlas.org/), \textit{(iii)} the tidal strength exerted by the first nearest neighbour, $Q_{\rm nn}$, from \citet{Argudo_Fernandez2015}, and \textit{(iv)} the weight for the volume correction, \textit{$V_{\rm max_w}$}, from the MaNGA Pipe3D value added catalog\footnote{Pipe3D value added catalog: \url{https://www.sdss4.org/dr17/manga/manga-data/manga-pipe3d-value-added-catalog/}}.

\section{Analysis}
\label{sec: Analysis}

\subsection{Sample Selection}

The original sample of this work comprises more than 10,000 galaxies of all morphological types, selected from the MaNGA targets in the final version of MaNGA Product Launch 11 (MPL-11), as part of SDSS Data Release 17 \citep[DR17,][]{Abdurrouf2022}. A subsample was selected based on the following criteria: (i) early-type galaxies were removed from the sample, (ii) edge-on galaxies with inclinations greater than 75° were excluded to minimize dust attenuation effects, (iii) nearly face-on galaxies with inclinations lower than 25° were removed to ensure unbiased estimation of rotational velocities, (iv) galaxies undergoing mergers were excluded based on their morphological distortions, and (v) only galaxies that have been carefully classified as barred and unbarred were included in the final sample. To identify barred galaxies, we make use of the morphological classification provided by Galaxy Zoo 2 \citep{Willett2013}, a citizen science project in which volunteers classified galaxies following a structured decision tree. To ensure a reliable identification of bars, we limited our analysis to galaxies with redshift $z < 0.06$. This choice follows the work of \citet{Hoyle2011}, who demonstrated that Galaxy Zoo 2 morphological classifications are robust within this redshift range. At higher redshift, the physical spatial resolution of SDSS imaging ($\sim 1.3 - 1.5''$) limits the ability to detect short bars. Consequently, shorter bars will tend to be missed or poorly resolved, as discussed in similar SDSS-based studies \citep[e.g.,][]{Gadotti2011,Erwin2018}. This selection effect implies that our bar fraction measurements are primarily representative of bars longer than this resolution limit, and it does not affect the comparative analysis of trends within our sample. For the bar classification, we required a minimum of 20 votes per task, and adopted thresholds of $p_{\rm disc} > 0.5$ and $p_{\rm{not\ edgeon}} > 0.5$ to select disc-dominated galaxies. A galaxy was considered barred if its bar vote fraction exceeded $p_{\rm{bar}} > 0.7$ (a selection similar to \cite{Cervantes-Sodi_Sanchez2017}), a conservative threshold designed to ensure a clean sample dominated by strongly barred systems \citep[see also][]{Masters2012, Kruk2018}. With this definition, weak bars are excluded from our analysis, minimizing contamination and ensuring that our results are based on robust bar identifications.

\subsection{Kinematic analysis}
\label{sec:Kin_ana}
For the selected subsample, we obtain the optimal kinematic data from the velocity maps following the procedure described in previous studies (e.g., \citet{Cortese2014ApJ}, \citet{Erik_Aquino2018,Erik_Aquino2020}). First, we exclude spaxels with velocity errors greater than 20 km/s, a threshold corresponding to one-third of the FWHM resolution of the MaNGA data ($\sim 70\ km/s$). Second, we retain only those galaxies for which at least $60\%$ of the spaxels within an ellipse, defined by a semi-major axis of $R_e$ or $2R_e$ and projected according to the position angle and inclination of each galaxy, satisfy this quality criterion. Following this procedure, our final sample comprises 1872 galaxies, 378 of which are barred and 1495 of which are unbarred.

The stellar velocity dispersion of each galaxy, $\sigma_{*}$, is calculated as the linear average of the velocity dispersions from all spaxels that satisfy our previously defined quality criteria. To determine the stellar rotation velocity, $V_{\rm rot}$, we generate a velocity histogram using all valid spaxels, which serves as a pseudo-integrated profile to characterize the system overall kinematic behavior. The width of this pseudo-integrated profile, denoted as $W$, is defined as the difference between the $10th$ and $90th$ percentile points of the velocity histogram: $W=V_{90}-V_{10}$. The observed width $W$ is corrected to the galaxy rest frame because cosmic expansion stretches velocity intervals by a factor of $(1+z)$. This standard “cosmological broadening” correction is applied by dividing the measured width by $1+z$ \citep[see e.g.,][]{Meyer2008,Catinella2012a,Catinella2012b,Barat2019}. We then combine this correction with the inclination deprojection to derive the intrinsic rotational velocity:

\begin{equation}
    V_{R} = \frac{W}{2(1+z)sin(i)},
\end{equation}
where $z$ is the redshift and $i$ is the galaxy inclination.

\subsubsection{Dynamical Mass Estimation at $R_e$ and $2Re$.}

To estimate the dynamical mass within one and two effective radii ($R_e$ and $2R_e$), we use the total kinematic parameter $S_K$ (defined below) together with the empirical calibration proposed by \citet{Erik_Aquino2018}:

\begin{equation}
    M_{\rm dyn} = \eta\frac{S_{K}^{2} R_{r}}{G},
    \label{ec. Mdyn}
\end{equation}
where $R_r$ is taken to be either the effective radius $R_e$ or twice the effective radius $2R_e$, depending on the aperture considered for the dynamical mass estimate, and $G$ is the gravitational constant. The calibration coefficient $\eta = 1.8$ is constant for all galaxies. This value originates from the empirical calibration, which is based on a suite of Jeans and Schwarzschild dynamical models constructed by \citet{Leung2018} under specific assumptions regarding galaxy shapes, projection effects, and dynamical structure. The factor $\eta$ does not encode the individual characteristics of each galaxy on a case-by-case basis; rather, it represents an average correction derived from those models and reflects the systematic effects of both our kinematic measurements and the adopted modeling assumptions, rather than galaxy-to-galaxy variations. The total kinematic parameter $S_K$ combines dynamical support from ordered rotation and random motions into a single quantity, providing a more robust tracer of the gravitational potential than either $V_{\rm rot}$ or $\sigma_{*}$ alone \citep[e.g.,][]{Weiner2006,Kassin2007}. It is defined as:

\begin{equation}
    S_K^2 = K V_{\rm rot}^2 + \sigma_{*}^2,
    \label{ec. Sk}
\end{equation}
where the stellar rotational velocity $V_{\rm rot}$ and velocity dispersion $\sigma_{*}$ are estimated as described in Sec. \ref{sec:Kin_ana}. To derive the value of the constant K, we known that for a spherically symmetric tracer population with isotropic velocity dispersion, a flat rotation curve, and a density profile $\rho \propto r^{-\alpha}$, one obtains $\sigma_{*} = V_{\rm rot}/\sqrt{\alpha}$ \citep[][2nd ed., Sec. 4.3]{Binney&Tremaine1987}, where $K \equiv 1/\alpha$. For disk galaxies with exponential stellar mass profiles, $\alpha \simeq 2-3$ \citep[][]{Weiner2006}, implying $K$ between $1/3$ and $1/2$. Following standard practice \citep[see e.g.,][]{Weiner2006,Kassin2007,Zaritsky2008,Covington2010,Kassin2012,Cortese2014,Simons2015,Straatman2017,Erik_Aquino2018,Erik_Aquino2020,Barat2019,Barat2020}, we adopt $K=1/2$ (hereafter $S_{K=0.5}$). Finally, because the calibration of Eq. \ref{ec. Mdyn} is not tailored to individual galaxies, it is important to verify that its use does not introduce systematic differences between barred and unbarred populations. To assess this, we performed an additional validation test using simulated galaxies from the highest–resolution run of the IllustrisTNG project, TNG50 \citep{Nelson2019TNG50, Pillepich2019TNG50}. By comparing the true dynamical masses of these simulated systems with the values recovered through the $S_{K=0.5}$–based estimator, we confirmed that the presence of a stellar bar does not introduce any systematic bias to the inferred dynamical mass within either $R_e$ or $2R_e$.

\subsubsection{Stellar angular momentum}

We estimate the stellar angular momentum of each galaxy using the
dimensionless parameter, $\lambda_{R}$, at a given radius, $R_e$ and $2R_e$, following the definition by \citet{Emsellem2007}:

\begin{equation}
    \lambda_{R} = \frac{\langle R|V| \rangle }{\langle R\sqrt{V^{2} + \sigma_{*}^2}\rangle} = \frac{\sum_{j=1}^{N_{spx}}F_{j}R_{j}|V_{j}|}{\sum_{j=1}^{N_{spx}}F_{j}R_{j}\sqrt{V_{j}^{2} + \sigma_{*,j}^{2}}},
\end{equation}
where the subscript $j$ marks each spaxel within an ellipse with a semi-major axis equal to the radius, $R_e$ or $2R_e$,  projected with each galaxy inclination, $i$, and position angle, $pa$, $F_j$ is the flux of the $j_{th}$ spaxel, $V_j$ and $\sigma_{*,j}$ are the stellar velocity and velocity dispersion, respectively. The quantity $\sqrt{V_j^2 + \sigma_{*,j}^2}$ in the denominator is proportional to the total mass (see eqs. \ref{ec. Mdyn} and \ref{ec. Sk}) and ensures that $\lambda_R$ is normalized, i.e., goes to unity when the mean stellar velocity dominates. This parameter $\lambda_{R}$ has become widely used as the projected stellar angular momentum of galaxies, particularly in studies employing integral field spectroscopy \citep[e.g.,][]{Emsellem2011,Fogarty2014,Seidel2015,Cortese2016,VandeSande2017,Veale2017b,Veale2017a,Graham2018,Falcon-Barroso2019,delMoral-Castro2020}. As $\lambda_{R}$ is derived from velocity fields subject to instrumental and seeing limitations, it must be corrected for observational effects such as beam smearing. The effect of seeing is to artificially smear the line-of-sight velocity field, thereby reducing the observed velocity gradients and increasing the measured velocity dispersion. As a result, the observed $\lambda_R$ is systematically biased toward lower values \citep[e.g.,][]{Graham2018,Harborne2019}. To address this, we apply the empirical correction proposed by \citet{Graham2018}, which models the seeing-induced bias in $\lambda_{R}$ based on the ratio of the PSF size to the galaxy effective radius and the intrinsic ellipticity. These corrections have been shown to recover intrinsic $\lambda_R$ values with reasonable accuracy for a broad range of galaxy morphologies and sizes. In our analysis, we adopt the corrected values of $\lambda_{Re}$ and $\lambda_{2Re}$ as the stellar angular momentum in the inner regions of galaxies.


\begin{figure*}
    \centering
    \begin{subfigure}[b]{0.33\textwidth}
        \includegraphics[width=\linewidth,height=6.0cm]{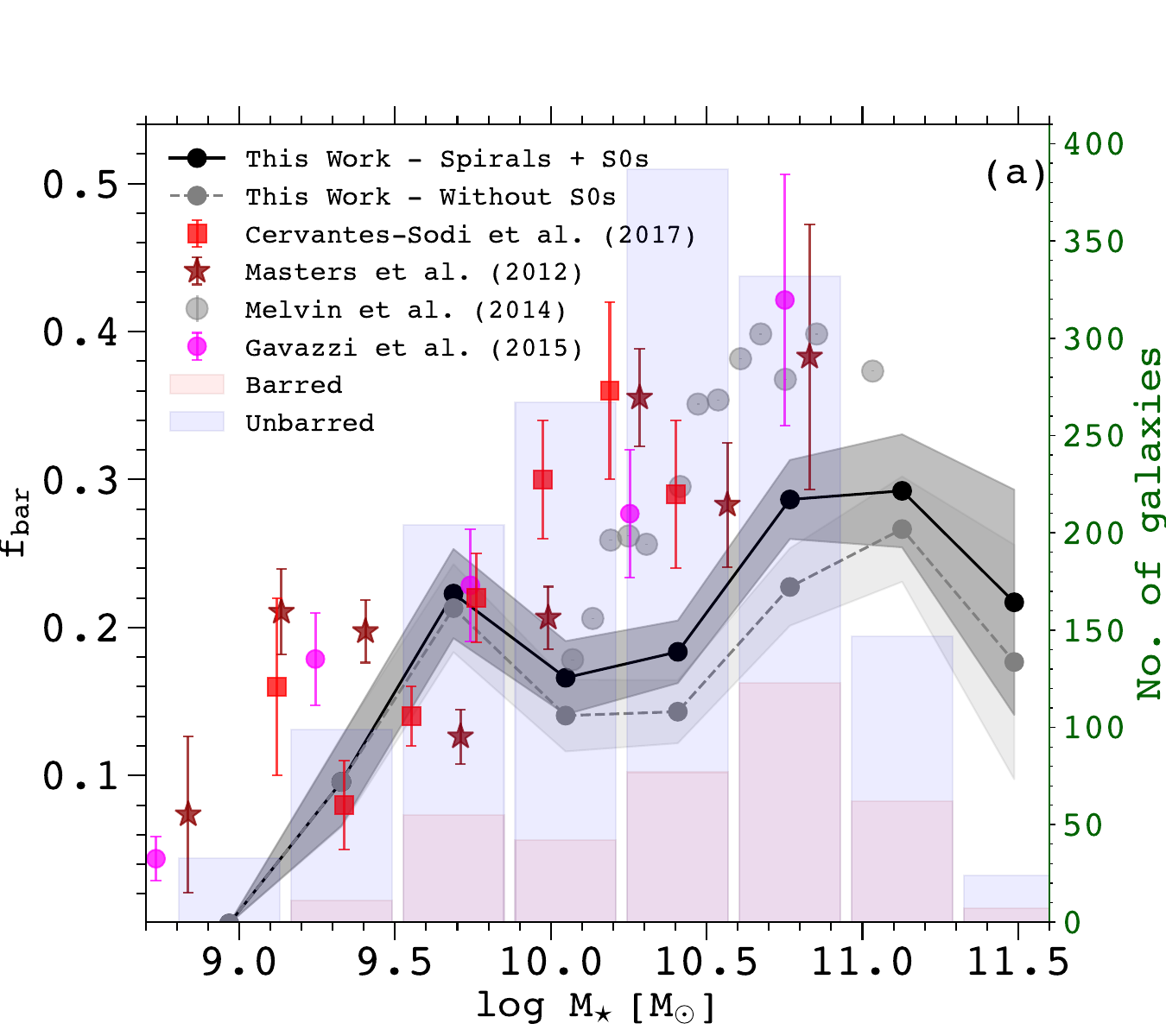}
        \caption[]{}
        \label{fig:fbar_mste}
    \end{subfigure}
    \hfill
    \begin{subfigure}[b]{0.33\textwidth}
        \includegraphics[width=\linewidth,height=6.0cm]{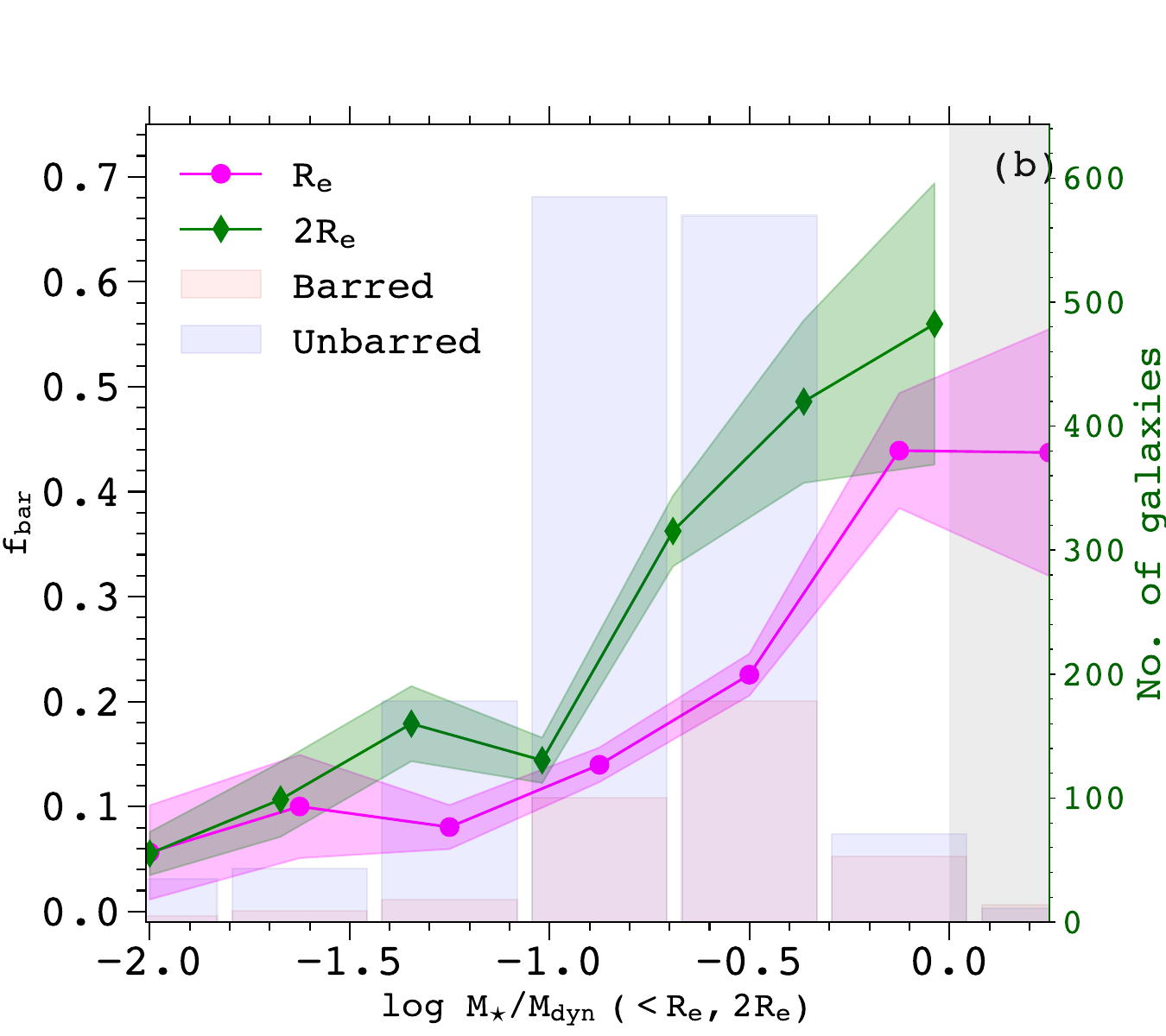}
        \caption[]{}
        \label{fig:fbar_MsteMdyn}
    \end{subfigure}
    \hfill
    \begin{subfigure}[b]{0.33\textwidth}
        \includegraphics[width=\linewidth,height=6.0cm]{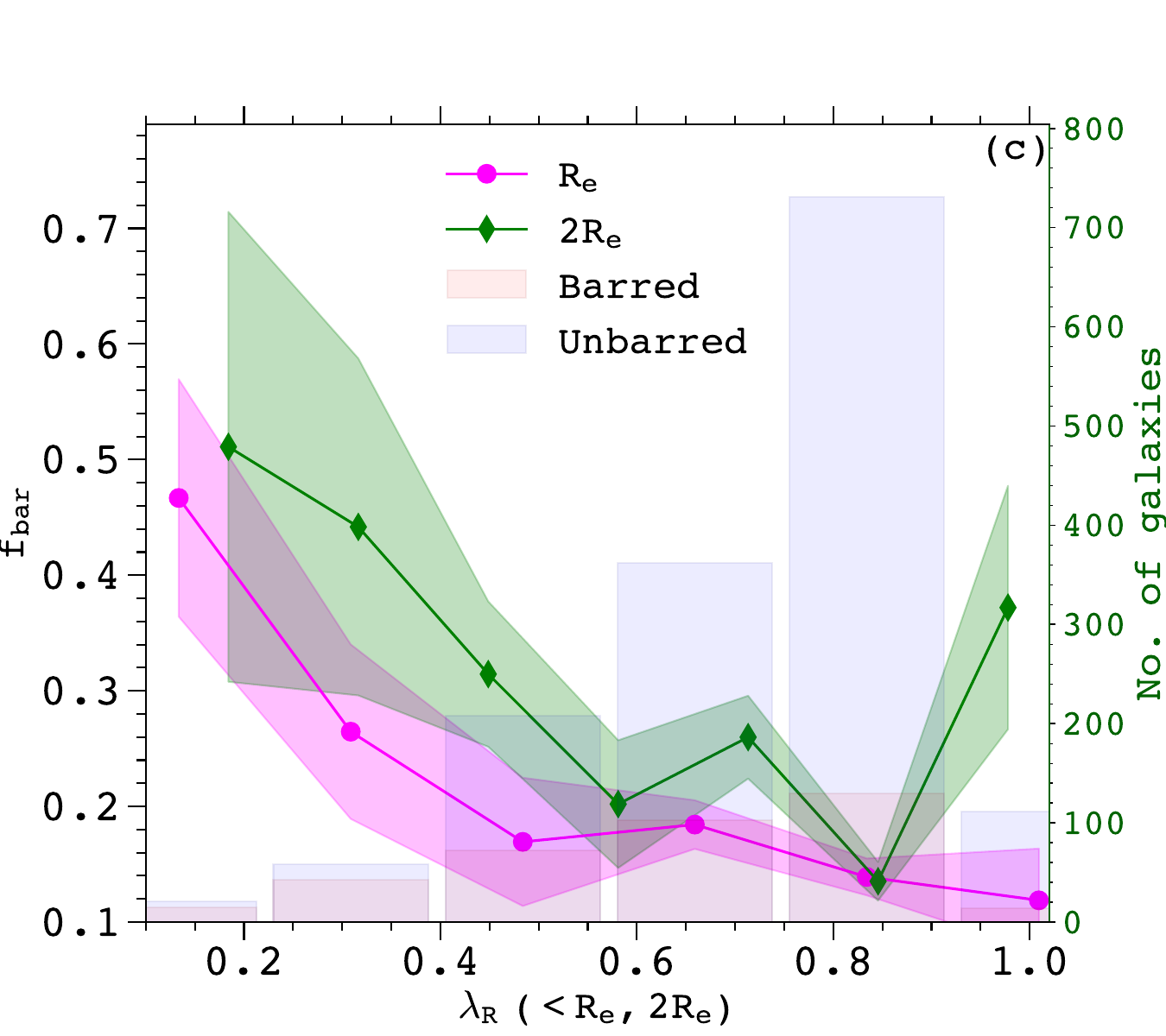}
        \caption[]{}
        \label{fig:fbar_lambda}
    \end{subfigure}
    \caption{Volume-weighted fraction of barred galaxies, $f_{\mathrm{bar}}$, as a function of: (a) total stellar mass, $\log M_{\star}$; (b) stellar-to-dynamical mass ratio, $M_{\star}/M_{\mathrm{dyn}}$ (measured within $R_e$ and $2R_e$); and (c) stellar angular momentum, $\lambda_R$ (also within $R_e$ and $2R_e$). The fraction is computed by binning the sample along the x–axis and taking the ratio of barred to total galaxies in each bin. Solid lines indicate the mean values in each panel, while the shaded regions correspond to the $1\sigma$ confidence intervals derived via bootstrap resampling. Blue (red) histogram displays the number of unbarred (barred) galaxies in each corresponding bin. On average, the bar fraction increases with stellar mass, and barred galaxies are more likely to be found in systems with high stellar mass dominance and low stellar angular momentum in their central regions.}
    \label{fig:fbar_all}
\end{figure*}

\section{Results and discussions}
\label{sec:Results}

\subsection{Dependence of the bar fraction on several galactic properties.}
\label{sec:bar_fraction}

We apply a volume correction to all estimates shown in the figures, using the $V_{\mathrm{max_w}}$ value from the PIPE3D VAC, which accounts for the maximum comoving volume over which each galaxy could be observed, given the survey flux limit. This weighting ensures that the statistical distributions and bar fractions represent the true underlying galaxy population rather than being biased toward brighter or more massive systems that are overrepresented in flux-limited samples. Bar fractions $f_{\rm bar}$ are computed by binning the sample along the x-axis variable and calculating the ratio of barred to total galaxies within each bin, therefore, represents an aggregate quantity (a fraction), not individual measurements. Figure \ref{fig:fbar_mste} shows the well-known trend between the bar fraction and the total stellar mass. We compare our results with previous studies \citep[e.g.,][]{Masters2012,Melvin2014, Gavazzi2015, Cervantes-Sodi2017}, which report a similar increase in bar fraction with increasing stellar mass. In our data, the bar fraction rises at low masses, reaching its first peak around $\log (M_\star/M_\odot) \sim 9.7$, then decreases near $\sim10^{10}\,M_\odot$, and subsequently increases again, attaining a second maximum between $\log (M_\star/M_\odot) \sim 10.7{-}11$. As noted by \citet{Erwin2018}, including S0 galaxies tends to flatten the high-mass tail of $f_{\rm bar}$. Our sample contains only a small number of S0 galaxies (N=67), unevenly distributed across the stellar–mass bins; when included, they produce only a slight increase in the high-mass end of the bar fraction. This suggests that the dip of $f_{\rm bar}$ around $\sim10^{10}\,M_\odot$ may be partly due to the lack of S0s in our dataset, preventing the flattening reported by \citet{Erwin2018}\, and partly an intrinsic trend within the spiral population.

Figure \ref{fig:fbar_MsteMdyn} shows the relationship between the bar fraction, $f_{\mathrm{bar}}$, and the stellar-to-dynamical mass ratio $\log(M_{\star}/M_{\mathrm{dyn}})$ at one and two effective radii ($R_e$). We find that bars are most frequent in systems with higher stellar-to-dynamical mass ratios, consistent with the idea that bars grow efficiently in massive auto-gravitating stellar discs. Observationally, this trend has been reported in studies showing that bar fraction decreases for systems with larger halo-to-stellar mass ratios \citep[e.g.,][]{Cervantes-Sodi2015,Díaz-Garcia2016A&A587A_160D,Cervantes-Sodi2017}, providing empirical support for the link between stellar dominance and bar formation. On the theoretical side, simulations have long predicted a stabilizing role of dark matter halos against bar formation \citep[e.g.,][]{Mayer2004}. More recently, hydrodynamical simulations by \citet{Reddish2022} show that bars rarely form in systems where dark matter dominates the inner galactic regions. Conversely, bars tend to form in stellar-dominated galaxies with higher stellar-to-dark matter ratios in their central parts, where the disc becomes more self-gravitating and conducive to bar formation \citep[e.g.,][]{Fujii2018,Fragkoudi2021,Bland-Hawthorn2023, RosasGuevara2024}. The observed trend suggests that galaxies with more stellar-dominated central regions (within $1R_e$) are more likely to host a bar. However, the relative role of bars compared to other evolutionary pathways, such as inside-out disc growth, minor mergers, or radial migration, remains to be fully disentangled \citep[see][for alternative mechanisms]{Roskar2015,Frankel2019}.

On the other hand, Figure \ref{fig:fbar_lambda} shows a clear anti-correlation between the bar fraction and the stellar angular momentum $\lambda_{R}$, measured at $R_e$ and $2R_e$. Although $\lambda_R$ is often used as a proxy for the galactic spin parameter \citep[][]{Harborne2019}, in our analysis, it is computed only for the stellar component within one or two effective radii. Therefore, the measured values are strongly affected by the presence of bars and by the lower angular momentum of the stars that make up the bar body. As a result, our $\lambda_R$ measurements primarily trace the kinematic imprint of bars rather than the intrinsic spin of the pre-bar disc. This interpretation is supported by the findings of \citet{Seidel2015}, who reported that the $\lambda_R$ profiles of barred galaxies present a dip in the inner regions, which they suggest is connected to the presence of inner substructures such as bars, highlighting how bars locally depress $\lambda_R$ values in the regions they dominate. Despite this caveat, the observed anti-correlation between bar fraction and $\lambda_R$ remains meaningful, as it illustrates how the presence of bars modifies the stellar angular momentum distribution in galaxies. Theoretical studies have long established the importance of angular momentum in disc stability and bar formation, showing that low-angular momentum discs are more susceptible to bar instabilities \citep{Peebles1969, Efstathiou1982, Athanassoula2003}, this behavior is consistent with the expectation that weakened rotational support facilitates the onset of disc instabilities that give rise to non-axisymmetric structures such as bars \citep[e.g.,][]{Lynden-Bell1972,Sellwood2014}. Moreover, simulations have shown that at high angular momentum, bars tend to dissolve or fail to form, resulting instead in weak oval distortions rather than prominent bar structures \citep[e.g.,][]{Long2014ApJ,Collier2018}. A concrete example of this behavior is found in \citet{Chim2025}, where the authors showed that the low bar fraction in LSB galaxies within the IllustrisTNG100 simulation is closely linked to their higher spin and gas-rich nature—supporting the idea that high angular momentum suppresses bar formation in these systems. From the observational point of view, \citet{CervantesSodi2013} found that galaxies with long bars tend to have the lowest spin values, while short-barred galaxies exhibit higher spin parameters, with the bar fraction peaking at intermediate values. \citet{Cervantes-Sodi_Sanchez2017} further showed that the lower bar fraction in low surface brightness (LSB) galaxies is primarily driven by their higher spin and gas content—two factors known to inhibit bar formation.

The results presented in this section indicate that barred galaxies are more frequently found in systems with higher stellar masses and lower specific angular momentum in their inner regions. Additionally, theoretical studies have established bars as efficient engines of redistributing angular momentum and gas inward within the corotation radius, leading to enhanced central mass concentrations \citep{Lynden-Bell1972, Sellwood1981, Athanassoula2003, Athanassoula2013, Kormendy2013}. Together, these insights are consistent with a scenario in which bars tend to appear in galaxies with particular internal conditions, like those observed in this section, and may subsequently influence their secular evolution. To further explore this possibility, the following section examines how the stellar-to-dynamical mass ratio and stellar angular momentum at one and two effective radii vary with total stellar mass, comparing barred and unbarred galaxies at fixed mass. This comparison highlights systematic differences between the two populations, which may be associated with the presence of bars and point toward distinct evolutionary pathways.

\begin{figure}
    \centering
    \includegraphics[width=1.0\linewidth,height=8cm]{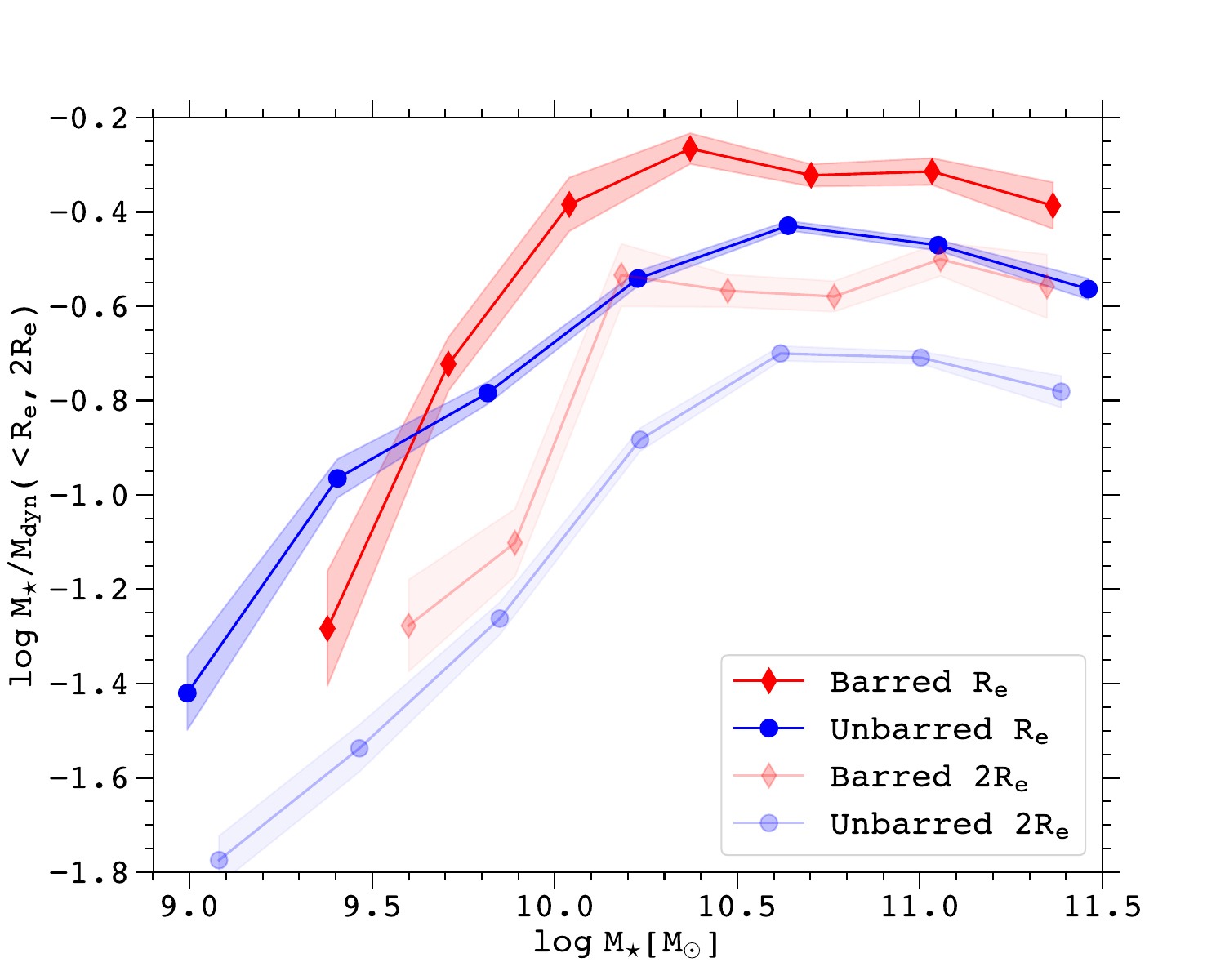}
    \caption{Volume-weighted stellar-to-dynamical mass ratio (measured within $R_e$ and $2R_e$) as a function of total stellar mass, $\log M_{\star}$, for barred (red lines and shaded regions) and unbarred (blue lines and shaded regions) galaxies. On average, barred galaxies exhibit a higher stellar mass dominance in their central regions at fixed total stellar mass than their unbarred counterparts.}
    \label{fig:MsteMdyn_Vs_Mstetot}
\end{figure}

\begin{figure}
    \centering
    {\includegraphics[width=1.0\linewidth,height=8cm]{Figs/lambda_Vs_totStemass_Vmax_w.pdf}}\\
    
    \caption{Volume-weighted stellar angular momentum, $\lambda_{R}$, (measured within $R_e$ and $2R_e$) as a function of the total stellar mass, $\log M_{\star}$, for barred (red lines and shaded regions) and unbarred (blue lines and shaded regions) galaxies. On average, above $\log M_{\star} \sim 9.7 M_{\odot}$, where the bar fraction significantly increases, barred galaxies exhibit lower stellar angular momentum than unbarred galaxies.}
    \label{fig:lambda_Vs_Mstetot}
\end{figure}

\subsection{Stellar Angular Momentum and Mass Dominance at Fixed Stellar Mass.}
\label{sec:lambda_ratio_Vs_stemass}

Given that stellar mass ($M_{\star}$) is a fundamental driver of galaxy evolution \citep[e.g.,][]{Kauffmann2003, Peng2010, Bluck2020}, we investigate how the stellar-to-dynamical mass ratio ($M_*/M_{\rm dyn}$) and stellar angular momentum ($\lambda_{R_e}$) vary at fixed stellar mass ($M_{\star}$) for barred and unbarred galaxies. Figure \ref{fig:MsteMdyn_Vs_Mstetot} shows the stellar-to-dynamical mass ratio as a function of the total stellar mass, where we observe a systematic offset between barred and unbarred galaxies. Barred galaxies exhibit higher stellar-to-dynamical mass ratios than unbarred galaxies at $R_e$ and $2R_e$, particularly at $ \log M_{\star} > 10^{9.75}\ M_{\odot}$. The trend is consistent with results from different cosmological simulations where barred galaxies are more frequently found in systems where stellar mass contributes more significantly within the inner regions; for example, in the TNG50 cosmological simulation, barred galaxies show enhanced stellar-to-dark matter mass ratios in their inner regions compared to unbarred systems \citep[e.g.,][]{Izquierdo-Villalba2022,Paula_D_Lopez2024}. Using the Auriga magneto-hydrodynamical cosmological zoom-in simulations, \citet{Fragkoudi2025} found that barred galaxies are typically dominated by stellar mass within their central five kpc over most of cosmic time, with this dominance becoming even stronger at the present epoch ($z\sim 0$). In contrast, unbarred galaxies tend to be more dark–matter dominated in their inner regions. From the observational point of view, \citet{Kashfi2023} using a sample of barred galaxies from the the Spitzer Photometry and Accurate Rotation Curves \citep[SPARC,][]{Lelli2016} database, show that practically all high-mass barred galaxies have maximal discs, i.e., stellar dominated galaxies where the disc contribution in the total rotation curve at $R \sim 2.2 R_d$ is $0.75 \leq V_{\rm disc}/V_{\rm tot} \leq 0.95$, where $R_d$ is the scale length of the disc. Similarly, \citet{Weiner2001} comparing the observed velocity field of the strongly barred galaxy NGC 4123 with gas-dynamical models show that their stellar disc dominates the rotation curve until well outside the optical radius of the galaxy. 

Regarding the kinematic properties, our analysis reveals that barred galaxies tend to have lower values of $\lambda_{R_e}$ and $\lambda_{2R_e}$ than unbarred galaxies of similar stellar mass (Figure~\ref{fig:lambda_Vs_Mstetot}). We emphasise that $\lambda_{R_e}$ is not a direct measure of the total angular momentum or spin of a galaxy, but rather an observational proxy of the stellar rotational support within a given aperture \citep{Emsellem2007, Emsellem2011}. The lower $\lambda_{R_e}$ observed in barred systems suggests that their inner regions are dynamically less rotationally supported than those of unbarred galaxies.
Such differences may arise from two complementary effects. First, stars in the bar region do not follow circular orbits but instead move along elongated trajectories aligned with the bar. As a consequence, the stellar orbits in these regions carry less specific angular momentum than those in the outer disc. Second, once a bar has formed, it can further redistribute angular momentum within the disc, transferring it from the inner regions to the outer disc and the dark matter halo \citep[e.g.,][]{Athanassoula2002, Debattista2006}. These effects reflect the kinematic imprint of the bar on the stellar component, rather than a direct tracer of the total angular momentum or spin of the system. These processes can gradually reduce the apparent rotational support in the central regions, thus lowering the observed $\lambda_{R_e}$.
Overall, the combined trends in $M_\star/M_{\rm dyn}$ and $\lambda_{R_e}$ suggest that bars are more likely to found in galaxies with a higher central stellar mass concentration and lower rotational support within their inner regions. These observational differences are consistent with a scenario in which both the initial internal structure and subsequent secular evolution shape the present-day properties of barred galaxies.

\begin{figure}
    \centering
    {\includegraphics[width=\linewidth,height=8cm]{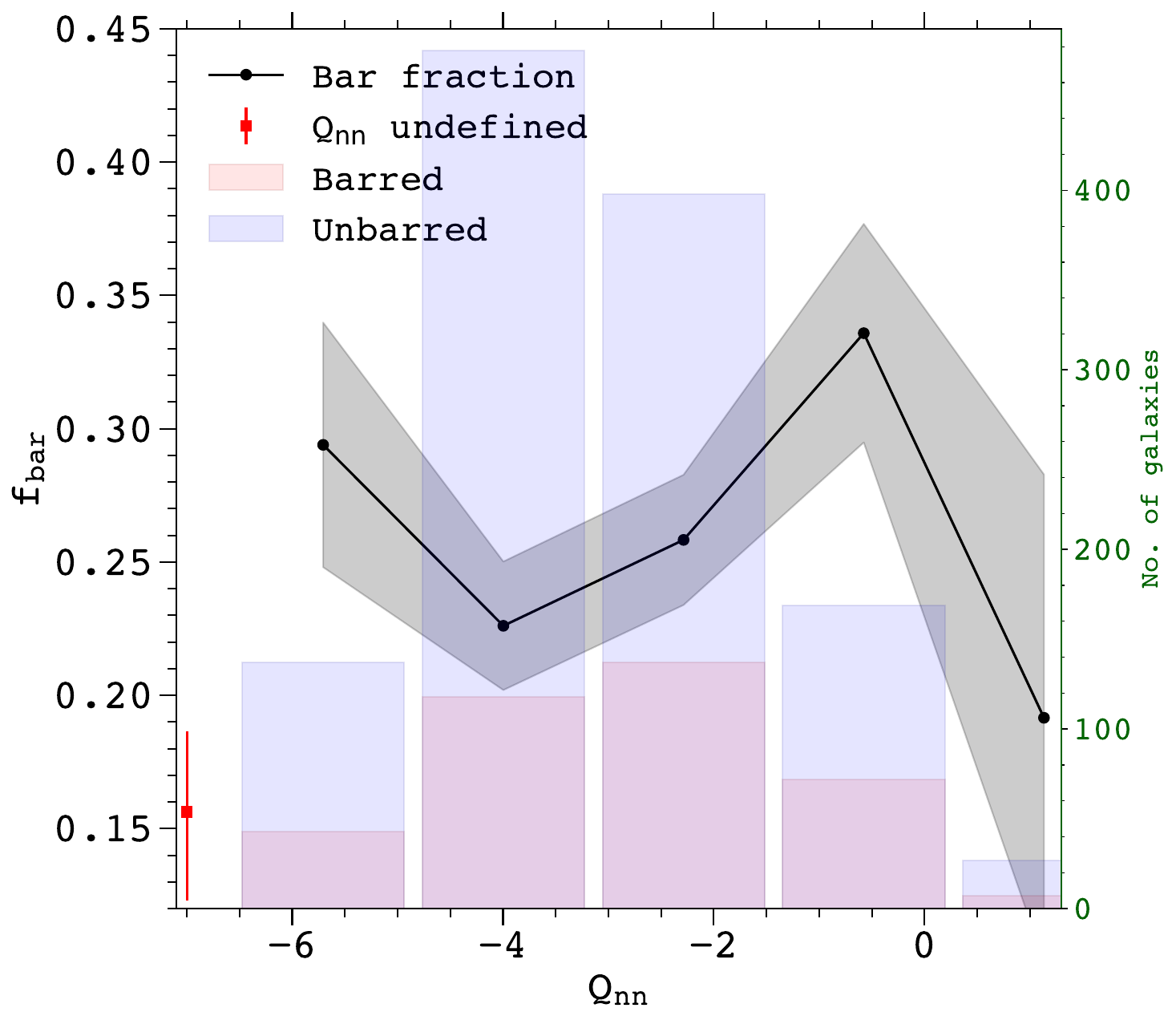}} \\

    \caption{Volume-weighted bar fraction, $f_{\mathrm{bar}}$, as a function of the tidal strength exerted by the first nearest neighbour, $Q_{\mathrm{nn}}$. Red (blue) histogram displays the number of barred (unbarred) galaxies in each $Q_{\rm nn}$ bin. The bar fraction shows a bimodal two-peak distribution, the first one at lowest $Q_{\mathrm{nn}}$ values, corresponding to galaxies minimally affected by external tidal perturbations (isolated galaxies), and the second one at high $Q_{\mathrm{nn}}$ values corresponding to tidal affected galaxies. The red point represents galaxies for which $Q_{\rm nn}$ could not be estimated due to the absence of a nearby neighbour. These cases are flagged as “NULL” in \citet{Argudo_Fernandez2015}, and we assign them a value of $Q_{\rm nn} = - 7$ for visualization purposes.}
    \label{fig:QLSS}
\end{figure}

\subsection{Bar Fraction as a Function of Tidal Strength Parameter.}
\label{sec:fbar_Vs_QLSS}

\begin{figure*}
    \centering
    \begin{subfigure}[b]{0.33\textwidth}
        \includegraphics[width=\linewidth,height=5.75cm]{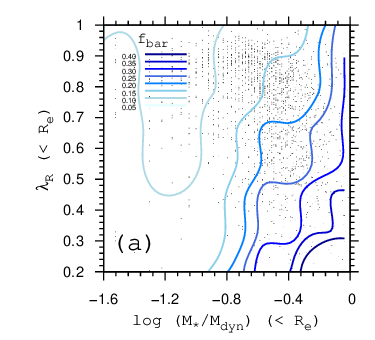}
        \caption[]{}
        \label{fig:Contorno_LamRatio}
    \end{subfigure}
    \hfill
    \begin{subfigure}[b]{0.33\textwidth}
        \includegraphics[width=\linewidth,height=5.75cm]{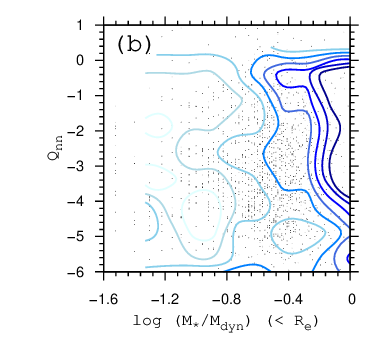}
        \caption[]{}
        \label{fig:Contorno_QlssRatio}
    \end{subfigure}
    \hfill
    \begin{subfigure}[b]{0.33\textwidth}
        \includegraphics[width=\linewidth,height=5.75cm]{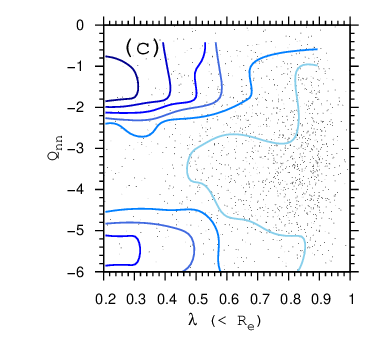}
        \caption[]{}
        \label{fig:Contorno_QlssLam}
    \end{subfigure}
    \caption{Bar fraction isocontours in key two-dimensional parameter spaces. All panels use volume-weighted statistics to account for survey incompleteness. Contours enclose $15-50\%$ of the sample, with colour gradients indicating the bar fraction. \textit{Panel (a)}: Stellar-to-dynamical mass ratio ($M_\star / M_\mathrm{dyn}$) versus specific angular momentum proxy ($\lambda_{R_e}$). Bar fraction increases toward systems that are simultaneously more stellar dominated and less rotationally supported, consistent with bar-driven secular evolution.
\textit{Panel (b)}: Stellar-to-dynamical mass ratio versus tidal strength parameter ($Q_{nn}$). Bar fraction peaks in galaxies with both high stellar dominance and moderate-to-high tidal strength, suggesting that interactions enhance bar formation in already bar-prone systems.
\textit{Panel (c)}: $\lambda_{R_e}$ versus $Q_{nn}$. Bar fraction declines at high $\lambda_{R_e}$, while at low $\lambda_{R_e}$ two populations emerge: one showing strong tidal interactions, and another more isolated. This suggests the existence of two bar formation pathways, one driven by intrinsic disc instabilities and another induced by the environment.
These multi-dimensional trends highlight that no single parameter governs bar formation; instead, bars emerge from a combination of internal dynamical conditions and external tidal influences.}
    \label{fig:contours_all}
\end{figure*}

It is well established through numerical simulations that external environmental influences can promote or delay bar formation \citep[e.g.,][]{MartinezValpuesta2017,Lokas2019b,Peschken2019,Semczuk2024,Zheng2025}. While secular processes dominate in isolated galaxies \citep[e.g.,][]{Kormendy2013}, tidal interactions can induce gravitational torques disrupting the disc stability even in galaxies that are stable against internal secular evolution and would otherwise remain unbarred in isolation \citep[e.g.,][]{Lokas2016, Peschken2019}. Although such external perturbations are not a necessary condition for bar formation \citep[][]{MartinezValpuesta2017}, they can significantly accelerate or modify the process in susceptible systems. To observationally assess the role of the environment in bar formation, Figure \ref{fig:QLSS} shows the bar fraction as a function of the tidal strength parameter of the nearest neighbor, $Q_{\rm nn}$. Galaxies with lower values of $Q_{\rm nn}$ correspond to systems that are less or not tidally affected (i.e. more isolated), whereas higher values indicate galaxies experiencing stronger tidal influences in denser environments. Galaxies with no identified neighbour are flagged as “NULL” in \citet{Argudo_Fernandez2015}; for visualization purposes, we assign them a nominal value of $Q_{\rm nn} = -7$ (shown as red points in Figure~\ref{fig:QLSS}). The distribution of the bar fraction, $f_{\rm bar}$, reveals a non-monotonic trend with two distinct peaks. The first occurs among isolated galaxies ($Q_{\rm nn} < -4$), where bars form efficiently in isolation unequivocally through internal secular processes. The second peak appears at moderate-to-strong tidal strengths ($-2 < Q_{\rm nn} < 0$), suggesting that interactions in denser environments can also foster bar formation. Conversely, $f_{\rm bar}$ decreases at the highest tidal strengths ($Q_{\rm nn} > 0$), possibly because intense gravitational encounters heat or distort disks, hindering the long-term survival of bars. These results suggest that while environmental factors can influence bar formation, they do not act as its primary driver. This interpretation is consistent with previous numerical and observational studies, which have shown that tidal interactions can either trigger or suppress bar formation, depending on galaxy properties and interaction strength \citep[e.g.,][]{Miwa1998, Lokas2016, Peschken2019}. For instance, \citet{Mendez-Abreu2012} proposed a bimodal environmental influence: interactions may promote bars in massive disks, while heating in low-mass systems inhibits their growth \citep[see also][]{Ghosh2021}. A similar mass dependence was reported by \citet{Aguerri2023}, who find that environmental effects can weaken bars in low-mass galaxies but have little impact on massive ones, emphasizing the dominant role of internal dynamics. Furthermore, the enhanced bar fraction we observe in denser environments ($Q_{\rm nn} > -2$) agrees with \citet{Skibba2012}, who found that group environments and minor mergers can induce disk instabilities that favor bar formation. Altogether, the significant fraction of bars observed in isolated systems and the environmental trends at higher densities support a hybrid scenario in which internal gravitational instabilities set the fundamental conditions for bar formation, while environmental interactions act as a secondary, modulating factor, either reinforcing or suppressing bars depending on galaxy mass and disk stability.


\subsection{Interplay Among Stellar Angular Momentum, Stellar-to-Dynamical Mass Ratio, and Environment in Barred Galaxies}

The high fraction of bars observed in massive galaxies with higher stellar-to-dynamical mass ratios, lower stellar angular momentum (Section \ref{sec:bar_fraction}) and higher values of tidal strength parameter ($Q_{\rm nn}$, Section \ref{sec:fbar_Vs_QLSS}) likely stems from a combination of internal secular evolution and external tidal triggering. While the environmental dependence of the bar fraction indicates that external processes can help trigger or accelerate bar formation, the observed trends in Section \ref{sec:lambda_ratio_Vs_stemass}, where barred galaxies exhibit higher stellar-to-dynamical mass ratios ($M_{\star}/M_{\rm dyn}$) and lower stellar angular momentum ($\lambda_{R_e}$) than unbarred ones, suggests that the presence of a bar and its subsequent-driven secular evolution leave a dominant imprint on galactic structure. However, the exact balance between these internal (secular) and external (tidal) mechanisms remains debated.

To directly disentangle the relative contributions of internal and external effects, we investigate in Figure \ref{fig:contours_all} the co-dependencies among $\lambda_{R_e}$, $\log(M_{\star}/M_{\rm dyn})$, the tidal strength exerted by the first nearest neighbour ($Q_{\rm nn}$), and the bar fraction ($f_{\rm bar}$). Figure \ref{fig:Contorno_LamRatio} shows the distribution of barred galaxies in the parameter space defined by the stellar angular momentum ($\lambda_{R_e}$) and the stellar-to-dynamical mass ratio ($M_{\star}/M_{\rm dyn}$) within one effective radius. The background contours indicate the bar fraction across this space.
 A clear gradient is observed: galaxies with higher $\lambda_{R_e}$ and lower $M_{\star}/M_{\rm dyn}$ exhibit the lowest bar fractions, whereas systems with lower $\lambda_{R_e}$ and higher $M_{\star}/M_{\rm dyn}$ show a markedly higher incidence of bars. This trend shows that bars preferentially occur in galaxies that are less rotationally supported and more centrally dominated by stellar mass. These internal properties may not only characterize systems that are more susceptible to bar instabilities but also reflect the subsequent secular evolution driven by the bars themselves. Indeed, bars are known to redistribute angular momentum, transferring it outward while driving gas toward the center \citep[e.g.,][]{Athanassoula1992, Sheth2005, Fragkoudi2016}. In addition, they can induce radial migration of stars through resonant interactions, thereby enhancing the central stellar concentration \citep[e.g.,][]{Sellwood&Binnet2002, Minchev2010, Fragkoudi2020}. Altogether, the observed gradient suggests a scenario in which internal dynamical conditions regulate the onset of bar formation, while secular processes driven by the bar itself further reshape the internal structure of galactic disks. 
 
 Figure \ref{fig:Contorno_QlssRatio} displays the bar fraction as a function of the tidal strength parameter $Q_{\rm {nn}}$ and the stellar-to-dynamical mass ratio within one effective radius. The contours reveal a structured trend in the incidence of bars across this parameter space. Galaxies with low $M_\star/M_{\rm dyn}$, where dark matter dominates the central potential, show the lowest bar fraction across the entire range of $Q_{\rm nn}$, suggesting that bars are less likely to form in dark matter-dominated systems in inner regions, regardless of whether they are isolated or in dense environments. In contrast, at high stellar-to-dynamical ratios, where the stellar component dominates the central potential, we find a substantial population of barred galaxies $(f_{\rm bar}\sim30)$ even at low $Q_{\rm nn}$ (isolated galaxies), and the fraction further increases $(f_{\rm bar}\sim 40)$ at high $Q_{\rm nn}$, where tidal interactions can trigger or enhance bar growth. Even at intermediate mass ratios $\log (M_\star/M_{\rm dyn}\sim -0.5)$, where the stellar component is not overwhelmingly dominant, the bar fraction rises at high $Q_{\rm nn}$, suggesting that tidal perturbations can induce bar formation in systems that would otherwise remain unbarred if left in isolation. These results indicate that a sufficient stellar-to-dynamical mass ratio is a necessary condition for bar formation, while environmental tides can act as an additional catalyst once the disc is near the instability threshold. 

Finally, Figure \ref{fig:Contorno_QlssLam} presents the bar fraction as a function of the tidal strength parameter $Q_{\rm {nn}}$ and the stellar angular momentum within one effective radius, $\lambda_{R_e}$. The bar fraction is encoded in the colored contours. The distribution is notably more scattered than previous parameter spaces, lacking a clear monotonic gradient. However, some trends persist: galaxies with high internal stellar angular momentum $\lambda_{R_e}$ show the lowest incidence of bars practically at all environments. In contrast, the most significant bar fractions occur at low $\lambda_{R_e} (< 0.6)$, where we observe a bimodality as a function of $Q_{\rm nn}$. This pattern suggests that, in isolated galaxies ($Q_{\rm nn} < -4$), low stellar angular momentum may favour bar formation and could also be influenced by the secular evolution driven by the bar itself. In contrast, for galaxies in denser environments $(-2 < Q_{\rm nn} < 0)$, tidal effects, while potentially important, play a secondary role in setting the overall bar fraction across this parameter space. Even, at intermediate and large stellar angular momentum $(\lambda_{R_{e}} > 0.6)$, the bar fraction rises at dense environments $(Q_{\rm nn}\sim -1)$, suggesting that tidal perturbations can induce bar formation in systems that would otherwise remain unbarred if left in isolation. For instance, galaxies with low stellar angular momentum but low $Q_{\rm nn}$ (isolated systems) evolved via secular processes, while those with low stellar angular momentum and high $Q_{\rm nn}$ have been influenced by external tides. Crucially, the parameter $Q_{\rm nn}$ introduces an environmental dimension to bar formation. The increasing bar fraction with $Q_{\rm nn}$ agrees with simulations where external perturbations trigger bar formation \citep[e.g.,][]{Lokas2016, MartinezValpuesta2017}. While the high bar fraction at lower $Q_{\rm nn}$ is consistent with the picture proposed by \citet{Li2023}, in which global gravitational instabilities predominantly produce bars in the local Universe, while external tidal perturbations play a secondary role, mainly facilitating bar growth in interacting systems when the internal gravitational instabilities are close, but not entirely sufficient, for spontaneous bar growth. The independent variations of $\lambda_{R_e}$, $\log(M_\star/M_{\rm dyn}$), and $Q_{\rm nn}$ indicate that no single parameter fully determines bar formation. This highlights the importance of considering both internal dynamics and environmental influences, as well as their interplay, to gain a deeper understanding of the diverse pathways that can lead to the formation and evolution of bars.

\section{Conclusions}
\label{sec:Conclusions}

Theoretically, bar formation in galaxies is generally attributed to internal dynamical instabilities, with several physical conditions determining the susceptibility of discs to develop bars. For example, baryon dominance, reflected in high stellar-to-dynamical mass ratios ($M_{\star}/M_{\mathrm{dyn}}$), which enhances disc self-gravity, and systems that are less rotationally supported (low stellar angular momentum, $\lambda_{R}$) both favor the onset of non-axisymmetric instabilities. The classical criterion of \citet{Efstathiou1982} encapsulates this intuition, however, more detailed theoretical work has shown that bar formation is also regulated by factors such as disc velocity dispersion, gas content, and the dynamical interplay with dark matter halos \citep[e.g.,][]{Athanassoula2003, Athanassoula2008, Sellwood2014}. Our results fit within this modern framework, supporting a scenario in which bars emerge preferentially in dynamically cold, baryon-dominated systems, with external tidal perturbations acting only as secondary modulators of bar growth, while strongest interactions appear to delay or suppress bar formation disrupting the conditions required for bar formation.

In summary, we confirm observationally that:

\begin{enumerate}
    \item Stellar bars are found preferentially in galaxies where the stellar component dominates over the dark matter halo within the central region (higher $M_{\star}/M_{\mathrm{dyn}}$). These barred galaxies show low stellar angular momentum $(\lambda_{R_e})$, which we interpret as a kinematic signature arising from the presence of the bar itself.

    \item At fixed stellar mass, barred galaxies exhibit higher stellar mass, within both one and two effective radii ($R_e$), and lower stellar angular momentum compared to unbarred galaxies. This difference becomes more pronounced for stellar masses above $\log(M_{\star}/M_{\odot}) \sim 9.7$, where the bar fraction also increases significantly.

    \item The bar fraction shows a clear bimodal dependence on the tidal strength parameter $Q_{\mathrm{nn}}$. Bars are more frequent both at low and high $Q_{\mathrm{nn}}$ values, suggesting two distinct regimes of bar formation: one associated with internally driven bars in relatively isolated systems, and another linked to externally induced bars in interacting environments.

    \item The bar fraction in interacting galaxies depends jointly on internal and external factors. We analyzed the parameter spaces by mapping bar fraction across the $M_{\star}/M_{\mathrm{dyn}}-\lambda_{R_e}$, $M_{\star}/M_{\mathrm{dyn}}-Q_{\mathrm{nn}}$, and $\lambda_{R_e}-Q_{\mathrm{nn}}$ planes. We find a clear gradient in bar fraction across the $M_{\star}/M_{\mathrm{dyn}}-\lambda_{R_e}$, confirming that these parameters are not independent; they jointly modulate bar formation and reflect the presence of a bar.

    \item Strong external tidal perturbations ($Q_{\rm nn} \sim -1$) appear to assist in triggering bar instabilities in galaxies that, under isolated conditions, would likely remain stable against bar formation. 

Our results point to a complex interplay between internal galaxy structure and environment in shaping bar formation. The evidence confirms that the internal gravitational instabilities are the primary driver, creating discs susceptible to bar formation. In this framework, external tidal perturbations play a secondary role: in very isolated galaxies, the absence of strong interactions allows internal secular processes to dominate, while in denser environments, tidal perturbations can trigger or enhance bar formation in predisposed systems. The observed decline in bar fraction at intermediate interaction strengths may reflect a transitional regime where moderate tidal heating temporarily stabilizes the disc against bar formation, whereas the strongest tidal interactions may suppress it.

\end{enumerate}

\section*{Acknowledgements}

EAO acknowledges  Universidad Nacional Autónoma de México Postdoctoral Program (POSDOC) for financial support. The authors acknowledge the financial support provided by PAPIIT project IN111825 from DGAPA-UNAM. We thank Daniel Jacobo Díaz-González for his valuable help in the optimization of specialized software. This research made use of Matplotlib \citep[][]{Hunter:2007}, SciPy \citep[][]{2020SciPy-NMeth}, Numpy \citep[][]{harris2020array} and Astropy \citep{astropy:2013, astropy:2018},\footnote{http://www.astropy.org} Python packages.


\section*{Data Availability}

The data underlying this article are publicly available from the following sources: (i) the volume correction weights, the line-of-sight kinematic maps, and the surface stellar mass density maps are included in the MaNGA Pipe3D Value Added Catalog, available at  \url{https://www.sdss4.org/dr17/manga/manga-data/manga-pipe3d-value-added-catalog/}; (ii) the effective radii and total stellar masses are obtained from the NASA-Sloan Atlas (NSA) catalog at \url{http://www.nsatlas.org/} (\citet{Blanton2011}); and (iii) the tidal strength values from the first nearest neighbor are taken from \citet{Argudo_Fernandez2015}. All data used in this study are publicly accessible through the referenced URLs or publications.




\bibliographystyle{mnras}
\bibliography{example} 








\bsp	
\label{lastpage}
\end{document}